\documentclass[a4paper,12pt]{article}
\usepackage[pctex32]{graphics}

\begin{document}
\topmargin 0pt \oddsidemargin 0mm
\newcommand{\beq}{\begin{equation}}
\newcommand{\eeq}{\end{equation}}
\newcommand{\beqa}{\begin{eqnarray}}
\newcommand{\eeqa}{\end{eqnarray}}
\newcommand{\fr}{\frac}
\renewcommand{\thefootnote}{\fnsymbol{footnote}}
\begin{titlepage}
\begin{flushright}
INJE-TP-04-04\\
astro-ph/0406387
\end{flushright}

\vspace{5mm}
\begin{center}
{\Large \bf Test of patch cosmology with WMAP} \vspace{12mm}

{\large  Kyong Hee Kim$^1$ and Yun Soo
Myung$^{1,2}$\footnote{e-mail
 address: ysmyung@physics.inje.ac.kr}}
 \\
\vspace{10mm} {\em  $^{1}$Relativity Research Center and School of
Computer Aided School, Inje University
Gimhae 621-749, Korea \\
$^{2}$Institute of Theoretical Science, University of Oregon,
Eugene, OR 97403-5203, USA}
\end{center}

\begin{abstract}
We calculate the  power spectrum, spectral index, and running
spectral index for inflationary patch cosmology arisen from
Gauss-Bonnet braneworld scenario using the Mukhanov equation. This
patch cosmology consists of  Gauss-Bonnet(GB), Randall-Sundrum
(RS-II), and  four dimensional (4D) cosmological models. There
exist several modifications in higher order calculations. However,
taking the power-law inflation by choosing different potentials
depending on the model, there exist minor changes up to second
order corrections. Since second order corrections are rather small
in the slow-roll limit, we could not choose a desired power-law
model which explains the WMAP data. Finally we discuss the
reliability of high order calculations based on the Mukhanov
equation by comparing the perturbed equation including 5D metric
perturbations. It turns out that  first order corrections are
reliable, while  second order corrections are not proved to be
reliable.
\end{abstract}
\end{titlepage}
\newpage
\setcounter{page}{2}
\section{Introduction}
There has been much interest in the phenomenon of localization of
gravity proposed by Randall and Sundrum\cite{RS2}. They assumed a
single positive tension 3-brane and a negative bulk cosmological
constant in the five dimensional (5D) spacetime. They obtained a
localized gravity on the brane by fine-tuning the tension  to the
cosmological constant. Recently, several authors studied
cosmological implications of brane world scenario. The brane
cosmology contains some important deviations from the
Friedmann-Robertson-Walker (FRW) cosmology\cite{BDL1,BDL2}. The
Friedmann equation is modified at high energy significantly.

On the other hand, it is generally accepted that curvature
perturbations produced during inflation are  the origin of
 anisotropies for CMB and inhomogeneities  for   galaxy formation and
other large-scale structures. The WMAP, SDSS and Lyman alpha  put
forward  constraints on cosmological models and confirm the
emerging standard model of cosmology, a flat $\Lambda$-dominated
universe seeded by scale-invariant adiabatic gaussian
fluctuations\cite{Wmap11,Wmap12,Wmap13,Wmap21,Wmap22}. In other
words, these results coincide with theoretical  predictions of the
slow-roll inflation based on general relativity and a single
inflaton. Further, the future experiments of Planck will be able
to place more stringent constraints on running spectral index than
those of WMAP and Lyman alpha.

If the brane inflation occurs, one expects that it gives us
different results in the high-energy regime.  Maartens et
al.\cite{MWBH} have described the inflationary perturbation in the
brane cosmology using the slow-roll approximation and potential
slow-roll parameters. Liddle and Taylor\cite{LT} have shown that
in the slow-roll approximation, the scalar perturbations alone
cannot be used to distinguish between the standard and brane
inflations.  Ramirez and Liddle\cite{RL} have studied the same
issue using the slow-roll approximation with Hubble slow-roll
parameters. They found that the
 first-order correction to the brane cosmology is of a similar size to that in
the standard cosmology. Also Tsujikawa and Liddle\cite{TL} have
investigated observational constraints on the  brane inflation
from CMB anisotropies by introducing the large-field, small-field,
and hybrid models. Unfortunately, in the slow-roll
approximation\cite{SL}, there is no significant change in the
power spectrum  between the standard and brane cosmology up to
first order corrections\cite{Cal1}. In order to distinguish
between the standard and brane inflations, it is necessary to
calculate their  power spectra up to second order in slow-roll
parameters using the slow-roll expansion\cite{SG}. Since second
order corrections are rather small in the slow-roll limit, it is
hard to discriminate between the standard and brane
inflations\cite{KLM1}.

Furthermore there exists the degeneracy between scalar and tensor
perturbations which is expressed as the consistency relation
$R=-8n_T$ in the standard inflation. This consistency relation
remains unchanged in the brane
cosmology\cite{HL1,HL2,HL3,Cal12,Cal13}. In order to resolve this
degeneracy problem, authors in\cite{DLMS} calculated the tensor
spectrum generated during inflation in the framework of the
Gauss-Bonnet braneworld. They found that this consistency relation
is broken by the Gauss-Bonnet term. However, this breaking of
degeneracy is ``mild" and thus the likelihood values are identical
to those in the standard and braneworld cases\cite{TSM}. Thus an
introduction of a Gauss-Bonnet term in the braneworld  could not
distinguish between the standard and brane inflations.

In the above approach, an important issue  to remark is that the
Mukhanov equation (\ref{eqsn}) was used for
 the study of 5D brane cosmology. Actually the Mukhanov equation
incorporates 4D metric (scalar) perturbations only and thus there
is no justification for using this to describe the effect of 5D
gravity on the brane. The 5D metric perturbations enter at first
order and  second order corrections to the power spectrum. Hence
one does not know whether or not  the Mukhanov equation is
reliable for studying the 5D brane cosmology. Recently, however,
Koyama and Soda\cite{KS} showed that on super-horizon scale, the
effect of 5D metric perturbations on the brane could be neglected
in comparison to 4D metric perturbations. Also Koyama, {\it et al}
\cite{KLMW} showed that even the effect of 5D metric perturbations
on the power spectrum appears to be large on  sub-horizon scale,
it is smaller than first-order corrections, irrespective of low
and high energies, on super-horizon scale. It turns out  that
 the Mukhanov equation  is valid for  the calculation of
cosmological parameters up to first order  because the
super-horizon perturbations during inflation are relevant to the
observation data.

 In this work, we will calculate the
power spectrum, spectral index, and running spectral index for
patch cosmology induced from the Gauss-Bonnet braneworld using the
slow-roll expansion. This cosmology consists of three regimes for
the dynamical history of the Gauss-Bonnet brane universe:
Gauss-Bonnet regime (GB), Randall-Sundrum brane cosmology in
high-energy regime (RS-II), and four dimensional cosmology (4D).
We follow notations of Ref.\cite{RL} except slow-roll
parameters\cite{SG}. Although second order corrections are too
small to be detected in current observations and their reliability
is not guaranteed, our work will provide a hint on explaining the
degeneracy between the standard and brane inflations.

The organization of our work is as follows. In Section II we
briefly review patch cosmology and slow-roll formalism. We
calculate relevant cosmological parameters of power spectrum,
spectral index, and running spectral index using the slow-roll
expansion in Section III. We choose  power-law  inflations to test
slow-roll inflation in patch cosmology and compare our results
with the  WMAP data in Section IV. In Section V we mention  the
consistency relation in patch cosmology. Finally we discuss our
results in Section VI. In Appendix A,  we derive the Mukhanov
equation including 5D metric perturbations from Koyama and Soda
expression and discuss the reliability of the Mukhanov equation
for higher order calculations. Explicit forms of potential
slow-roll parameters are shown in Appendix B for patch cosmology.

\section{Patch cosmology}

We start with the two Friedmann equations arisen from Gauss-Bonnet
brane cosmology by adopting a flat Friedmann-Robertson-Walker
(FRW) metric as the background spacetime on the
brane\cite{Cal1,DLMS,TSM}
\begin{equation}
\label{Heq} H^2=\beta^{2}_{q}\rho^{q},~\frac{\dot{H}}{\beta_q}=
-\frac{3q\beta_q}{2}\Big(\frac{H}{\beta_q}\Big)^\theta(\rho+p)
\end{equation}
where $H=\dot{a}/a$, $q$ is a parameter labelling  a model, and
$\beta_{q}^{2}$ is a factor with energy dimension
$[\beta_q]=E^{1-2q}$.
 An additional  parameter
$\theta=2(1-1/q)$ is introduced for our purpose. In deriving the
latter equation, one uses the continuity equation of
$\dot{\rho}+3H(\rho+p)=0$. In this work, we neglect a holographic
term from Weyl tensor because its form of $1/a^4$ decreases rather
than a curvature term of $k/a^2$ during inflation, and the
bulk-brane exchange because we don't know  yet how to accommodate
its explicit form to the Friedmann equation on the
brane\cite{holo1,holo2,holo3}. We call the above defined on
$q$-dependent energy regimes as a whole ``patch cosmology". We
summarize relevant models and their parameters  in patch
cosmology: 1) for GB, $q=2/3 (\theta=-1),\beta^2_{2/3}=
(\kappa^2_5/16\tilde{\alpha})^{2/3}$.  2) for 4D, $q=1
(\theta=0),\beta^2_1= \kappa_4^2/3$. 3) for RS-II, $q=2
(\theta=1),\beta^2_2=\kappa^2_4/6\lambda$. $\kappa^{2}_{5}=8\pi
G_5$ is the 5D gravitational coupling and $\kappa^2_4=8\pi G_4$ is
the four-dimensional gravitational coupling.
$\tilde{\alpha}=1/8g_s$ is the Gauss-Bonnet coupling, where $g_s$
is the string energy scale, and $\lambda$ is the RS brane tension.
A relation between these is $\kappa^2_4/\kappa^2_5=\mu/(1+\beta)$,
where $\beta=4\tilde{\alpha}\mu^2,~\mu=1/\ell$ with  AdS$_5$
curvature radius $\ell$. RS-II case of $\mu=\kappa_4^2/\kappa_5^2$
is recovered when $\alpha=0$.
\begin{figure}[t!]
\includegraphics{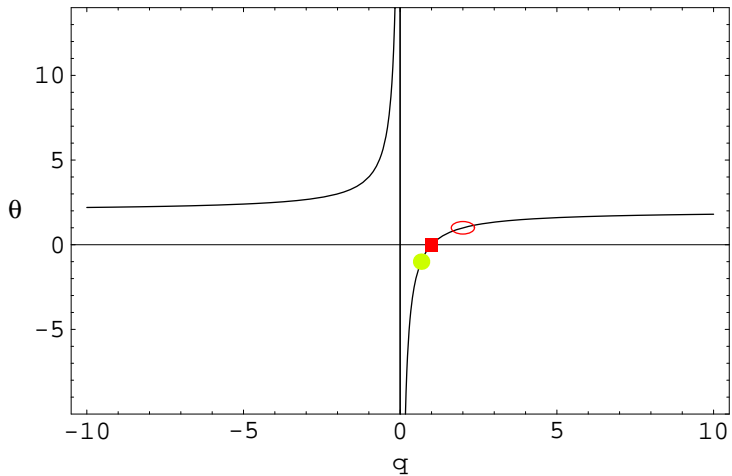} \caption{A graph for the
parameter $\theta(q)$. Three models are located at
GB($\theta(2/3)=-1$), 4D($\theta(1)=0$), and RS-II($\theta(2)=1$),
respectively.} \label{fig1}
\end{figure}

In the
 Gauss-Bonnet high-energy regime of $\sigma/\sigma_0 \gg 1$ with
the matter energy density $\sigma=\rho+\lambda$ and
$1/\sigma_0=\sqrt{\alpha/2}\kappa^2_5$, we have a non-standard
cosmology  called ``GB" model. When the energy density is far
below the 5D/string scale ($\lambda/\sigma_0 \ll \sigma/\sigma_0
\ll 1$) but $\rho \gg \lambda$, we have the brane cosmology in
high-energy regime called as ``RS-II" model. The four-dimensional
cosmology(``4D") is recovered when $\rho/\sigma_0 \ll
\sigma/\sigma_0 \ll 1$ but with $\rho \ll \lambda$.  A plot for
$\theta(q)$ is shown in Fig.1.  We wish to comment on the two
limiting cases of $q\to 0$ and $q \to \infty$. In the case of
$q\to0~(\theta\to \infty)$, one recovers de Sitter spacetime when
$\beta^2_0 \propto \Lambda$, whereas in the case of $q \to
\infty~(\theta=2)$, one finds an interesting case in the power-law
inflation.

 Introducing an inflaton $\phi$ confined to the
brane, one finds the equation
\begin{equation}
\label{seq} \ddot{\phi}+3H\dot{\phi}=-V^{\prime},
\end{equation}
where dot and prime denote the  derivative with respect to time
and $\phi$, respectively. Its energy density and pressure are
given by $\rho=\dot{\phi}/2+V$ and $p=\dot{\phi}/2-V$. From now on
we use the slow-roll formalism for inflation: an accelerated
universe $(\ddot a>0)$ is driven by a single scalar field slowly
rolling down its potential toward a local minimum. This means that
Eqs.(\ref{Heq}) and (\ref{seq}) take the following form
approximately:
\begin{equation}
H^2\approx \beta^2_q V^q,~ \dot{\phi}\approx -V'/3H.
\end{equation}

 In order to take this approximation into account, we
introduce Hubble slow-roll parameters (called H-SR towers) on the
brane as
\begin{equation} \label{slpa}
\epsilon_1 \equiv -\frac{\dot H}{H^2}\approx
\fr{3q\beta^{2-\theta}_q}{2}\fr{\dot{\phi}^2}{H^{2-\theta}},~~\delta_n
\equiv \frac{1}{H^n\dot{\phi}}\frac{d^{n+1}\phi}{dt^{n+1}}
\end{equation}
which satisfy the slow-roll condition:
$\epsilon_1<\xi,~|\delta_n|<\xi^n$ for some small perturbation
parameter $\xi$ defined on the brane.
 Here
 the subscript denotes  slow-roll (SR)-order in the slow-roll
expansion. We note that the original definition of H-SR parameters
is independent of $q$ because these are constructed in a geometric
way.

\section{Cosmological parameter calculation}
We are now in a position to calculate cosmological parameters
using the Mukhanov's formalism for scalar perturbations. We
introduce a new variable $u^q\equiv a Q^q= a(\delta
\phi^q-\dot{\phi}\psi/H)$ where $ \delta \phi^q$ is a perturbed
inflaton. $\psi$ is a perturbed metric function defined in
$ds^2=-(1+2A)dt^2+(1+2\psi)a^2 d {\bf x} \cdot  d {\bf x}$. Its
Fourier modes $u^q_k$ in the linear perturbation theory satisfies
the Mukhanov equation:
\begin{equation}
\label{eqsn} \frac{d^2u^q_k}{d\tau^2} +
\left(k^2-\frac{1}{z}\frac{d^2z}{d\tau^2}\right)u^q_k =
0,\end{equation} where the $q$-dependent potential-like term is
given by \footnote{Here one change in coefficient of
$\epsilon_1^2$ occurs : $1 \to \frac{1}{q}$. Although the full
Gauss-Bonnet brane cosmology provides a complicated potential-like
term, its patch approximation provides the Mukhanov equation with
the nearly same potential-like terms except one term of
$\epsilon_1^2$. This is why we choose patch cosmology instead of
the Gauss-Bonnet brane cosmology in the beginning. Thanks to a
minor change, one expects to find the same cosmological parameters
when working the slow-roll approximation with the first-three
terms in the potential-like term.}
\begin{equation} \label{z-pot}
\frac{1}{z}\frac{d^2z}{d\tau^2}= 2a^2H^2\Big(
1+\epsilon_1+\frac{3}{2}\delta_1+ \frac{1}{q}\epsilon_1^2
+2\epsilon_1\delta_1 +\frac{1}{2}\delta_2\Big).
\end{equation}
Here $\tau$ is the conformal time defined by $d\tau=dt/a$, and
$z=a\dot{\phi}/H$ encodes all information about a slow-roll
inflation with $\dot{\phi}=\sqrt{\rho+p}$.

 Before we proceed, we
have to mention that Eq.(\ref{eqsn}) is the nearly same form as in
the conventional 4D perturbation theory\cite{muk1,muk2,muk3}. It
is well known that the perturbation theory of braneworlds
including Randall-Sundrum and Gauss-Bonnet models is very
different from the 4D perturbation
theory\cite{MWBH,LT,RL,TL,HL1,HL2,HL3,DLMS,LMW}. Making use of the
4D Mukhanov equation to study the braneworld perturbation, the
problem  is that this equation incorporates  4D metric (scalar)
perturbations only and thus there is no justification for using
this to describe the effect of  5D gravity on the brane. This
falls short of being a full 5D calculation as is required by the
braneworld scenario. The 5D metric perturbations entered at first
order and second order corrections to the perturbed
equation\cite{KS}. Therefore it is not evident that the Mukhanov
equation is reliable for studying the 5D brane cosmology. However,
it was shown recently that even though the effect of 5D metric
perturbations on inflation appears to be large on small-scales
(sub-horizon), on large-scales (super-horizon) this effect is
smaller certainly than first order corrections to de Sitter
background\cite{KLMW}. Further, the effect of 5D metric
perturbations is very small, at low energies, on super-horizon and
also this is suppressed, even at high energies, on super-horizon.
In Appendix A, we derive the Mukhanov equation including 5D metric
perturbations from Koyama and Soda expression (Eq.(C.5) in
Ref.\cite{KS}). Therefore it is sensible to use the Mukhanov
equation (\ref{eqsn}) to compute first order corrections to
cosmological parameters on the super-horizon scale.

 In general its
asymptotic solutions are obtained as
\begin{equation}\label{bc}
u^q_k \longrightarrow \left\{
\begin{array}{l l l}
\frac{1}{\sqrt{2k}}e^{-ik\tau} & \mbox{as} & -k\tau \rightarrow \infty \\
{\cal C}^q_k z & \mbox{as} & -k\tau \rightarrow 0.
\end{array} \right.
\end{equation}
 The first solution corresponds to a plane wave on
scale much smaller than the Hubble horizon of $d_H=1/H$
(sub-horizon regime), while the second is a growing mode on scale
much larger than the Hubble horizon (super-horizon regime). Using
a relation of  $R^q_{c{\bf k}}=-u^q_{\bf k}/z$ with $u^q_{\bf
k}(\tau)=a_{\bf k}u^q_k(\tau)+a^{\dagger}_{-{\bf
k}}u^{q*}_k(\tau)$ and a definition of
$P^q_{R_c}(k)\delta^{(3)}({\bf k}-{\bf
l})=\fr{k^3}{2\pi^2}<R^q_{c{\bf k}}(\tau)R^{q \dagger}_{c{\bf
l}}(\tau)>$, one finds the power spectrum for a curvature
perturbation in the super-horizon regime
\begin{equation}\label{gps}
P^q_{R_c}(k) = \left(\frac{k^3}{2\pi^2}\right)
\lim_{-k\tau\rightarrow0}\left|\frac{u^q_k}{z}\right|^2 =
\frac{k^3}{2\pi^2}|{\cal C}^q_k|^2.
\end{equation}
Our task is to find ${\cal C}^q_k$ by solving the Mukhanov
equation (\ref{eqsn}). In general it is hard to solve this
equation. However, we can solve it using either the slow-roll
approximation \cite{SL} or the slow-roll expansion\cite{SG}.
 In the slow-roll approximation  we take  $\epsilon_1$ and $\delta_1$ to be constant.
Thus this method  could not be considered as a general approach
beyond the first-order correction to the power
spectrum\cite{KLM,KLLM}. In order to show different power spectra
depending on $q$, one uses the slow-roll expansion based on
Green's function technique. A step  to consider is a slowly
varying nature of slow-roll parameters implied by
$\dot{H}=-\frac{3}{2}q\beta^2_q \rho^{q-1}(\dot{\phi})^2$ and
$\ddot{H}=2H\dot{H}[-(1-1/q)\epsilon_1 +\delta_1]$:
\begin{eqnarray} \label{slow-di}
\dot \epsilon_1
&=&2H\Big(\epsilon_1^2/q+\epsilon_1\delta_1\Big),~\dot{\delta}_1=
H(\epsilon_1\delta_1-\delta^2_1+\delta_2),\\
\dot{\delta_2}&=&H(2\epsilon_1\delta_2-\delta_1\delta_2+\delta_3),~
\dot{\delta_3}=H(3\epsilon_1\delta_3-\delta_1\delta_3+\delta_4)
\end{eqnarray}
which means that derivative of slow-roll parameters with respect
to time increases their SR order by one in the slow-roll
expansion. Note that except $\dot{\epsilon_1}$, all of
$\dot{\delta}_n$ are independent of $q$. In this sense our choice
for H-SR towers is convenient to investigate  patch cosmology in
compared with others in Ref.\cite{STG,Cal1}. After a lengthly
calculation following ref.\cite{SG}, we find the $q$-power
spectrum
\begin{eqnarray}
 \label{2ndps}
P^{q}_{R_c}(k) & = &  \fr{H^4}{(2\pi)^2\dot{\phi}^2} \left\{ 1
-2\epsilon_1 + 2\alpha(2\epsilon_1+\delta_1) \right. \\
&& \left.
+\left((8-4/q)\alpha^2-4(1-1/q)\alpha-(19+4/q)+(2+1/3q)\pi^2\right)\epsilon_1^2 \right. \nonumber \\
&& \left. +
\left(3\alpha^2+2\alpha-22+29\pi^2/12\right)\epsilon_1\delta_1 +
\left(3\alpha^2-4+5\pi^2/12\right)\delta_1^2 +
\left(-\alpha^2+\pi^2/12\right)\delta_2 \right\} \nonumber
\end{eqnarray}
and the right hand side should be evaluated at horizon crossing of
$k=aH$. $\alpha$ is defined by $\alpha=2-\ln2-\gamma \simeq
0.7296$ where $\gamma$ is the Euler-Mascheroni constant, $\gamma
\simeq 0.5772.$ We note that $q-$dependent terms appear only in
coefficient of $\epsilon_1^2$.  Using
 $\fr{d\epsilon_1}{d\ln k}=
2(\epsilon_1^2/q+\epsilon_1\delta_1)/(1-\epsilon_1),~\fr{d\delta_1}{d\ln
k}=(\epsilon_1\delta_1-\delta_1^2+\delta_2)/(1-\epsilon_1),~\fr{d\delta_2}{d\ln
k}=(2\epsilon_1\delta_2-\delta_1\delta_2+\delta_3)/(1-\epsilon_1)$,
and $\fr{d\delta_3}{d\ln
k}=(3\epsilon_1\delta_3-\delta_1\delta_3+\delta_4)/(1-\epsilon_1)$,
the $q$-spectral index defined by
\begin{equation}
n_{s}^q(k) = 1 + \frac{d \ln P^{q}_{R_c}}{d \ln k}
\end{equation}
can be calculated up third order

\begin{eqnarray}
\label{11} n^{q}_{s}(k) =& & 1 - 4\epsilon_1 - 2\delta_1
+(-4-4/q+8\alpha/q)\epsilon_1^2
              + (10\alpha -6)\epsilon_1\delta_1 -2\alpha\delta_1^2+2\alpha\delta_2
              \\ \nonumber
          & & + \left(
                    -16\alpha^2/q^2+(16/q^2+24/q)\alpha-4-16/q^2-88/q +
                    (4/3q^2+8/q)\pi^2
                \right)\epsilon_1^3       \\ \nonumber
          & &     +\left(
                    -(26/q+5)\alpha^2+(32+28/q)\alpha-112-60/q+(125/12+37/6q)\pi^2
                \right)\epsilon_1^2\delta_1
                \\ \nonumber
          & & + \left(
                    -3\alpha^2+4\alpha-30+13\pi^2/ 4
                \right)\epsilon_1\delta_1^2
                +\left(
                    -7\alpha^2+8\alpha-22+31\pi^2/12
                \right)\epsilon_1\delta_2
                \\ \nonumber
          & & + \left(
                    -2\alpha^2+8-5\pi^2/6
                \right)\delta_1^3
                +\left(
                    3\alpha^2-8+3\pi^2/4
                \right)\delta_1\delta_2
                +\left(
                    -\alpha^2+\pi^2/12
                \right)\delta_3.
\end{eqnarray}
Here we find three changes in $\epsilon_1^2,~\epsilon^3_1$ and
$\epsilon^2_1 \delta_1$.
 Finally the $q$-running
spectral index up to fourth order is determined  by

\begin{eqnarray}
\label{12} \frac{d}{d\ln k} n^{q}_{s} & = &
-8\epsilon^{2}_{1}/q-10\epsilon_1\delta_1+2\delta^{2}_{1}-2\delta_2
\\ \nonumber
&+& \left(-16/q^2-24/q+32\alpha/q^2\right)\epsilon^{3}_{1}
\\ \nonumber
&+& \left(-32 -28/q+(10+52/q)\alpha\right)\epsilon^{2}_{1}\delta_1
\\ \nonumber
&+&(6\alpha-4)\epsilon_1\delta^{2}_{1}+(14\alpha
-8)\epsilon_1\delta_2+4\alpha\delta^{3}_{1}-6\alpha\delta_1\delta_2+2\alpha\delta_3 \\
\nonumber
         &+&\left(-96\alpha^2/q^2+(96/q^3+176/q^2)\alpha-(96/q^3+544/q^2+48/q) +
         (8/q^3+48/q^2)\pi^2\right)\epsilon_1^4
         \\ \nonumber
         &+& \left(-(200/q^2+46/q+5)\alpha^2+(208/q^2+352/q+42)\alpha \right)\epsilon_1^3\delta_1 \\ \nonumber
         &+&\left(-(336/q^2+1064/q +168)+(98/3q^2+575/6q+125/12)\pi^2\right)\epsilon_1^3\delta_1
               \\ \nonumber
         &+&\left(-(84/q+21)\alpha^2+(92/q+100)\alpha-(240/q+400)+(25/q+151/4)\pi^2
             \right)\epsilon_1^2\delta_1^2  \\ \nonumber
         &+&\left(-(40/q+19)\alpha^2+(44/q+62)\alpha-(104/q+164)+(34/3q+187/12)\pi^2
         \right)\epsilon_1^2\delta_2
               \\ \nonumber
         &+&\left(-6\alpha^2+4\alpha+24-5\pi^2/
             2\right)\epsilon_1\delta_1^3
            +\left(-4\alpha^2+10\alpha-106+34\pi^2/
             3\right)\epsilon_1\delta_1\delta_2
               \\ \nonumber
         &+&\left(-10\alpha^2+10\alpha-22+17\pi^2/
             6\right)\epsilon_1\delta_3
            +\left(6\alpha^2-24+5\pi^2/2\right)\delta_1^4
               \\ \nonumber
         &+&\left(-12\alpha^2+40-4\pi^2\right)\delta_1^2\delta_2
            +\left(4\alpha^2-8+2\pi^2/3\right)\delta_1\delta_3
               \\ \nonumber
         &+&\left(3\alpha^2-8+3\pi^2/4\right)\delta_2^2
            +\left(-\alpha^2+\pi^2/12\right)\delta_4.
            \nonumber
\end{eqnarray}
Here we have several changes in
$\epsilon_1^2,~\epsilon_1^3,~\epsilon_1^2
\delta_1,~\epsilon_1^4,~\epsilon_1^3 \delta_1,~\epsilon_1^2
\delta_1^2,~\epsilon_1^2 \delta_2$. Up to now we calculate the
power spectrum, spectral index, and running spectral index for
slow-roll inflations in patch cosmology.  If one uses the
slow-roll approximation,  there is no apparent distinction in
power spectrum between GB, 4D, and RS-II. However, as are shown in
Eqs.(\ref{2ndps}), (\ref{11}), and (\ref{12}), several
modifications appear in the higher-order corrections. This is our
motivation of why to calculate up to higher-order corrections
using the slow-roll expansion. That is, we need to know the
apparent distinction between GB, 4D, and RS-II (three models of
patch cosmology) when applying them to describe the inflationary
perturbations.

We note here that first order calculations in Eq.(\ref{2ndps}),
second order calculations in Eq.(\ref{11}), and third order
calculations in Eq.(\ref{12}) are only reliable if one takes into
5D metric perturbations account seriouly.

 \section{Power-law inflation}

 As a concrete
example, we choose  the power-law inflation like $a(t)\sim t^p$ to
test patch cosmology. Although second order corrections are very
small in the slow-roll limit and their reliability is not
guaranteed, we calculate cosmological parameters up to second
order to understand a degeneracy between the standard and brane
inflations. Then Hubble slow-roll parameters (H-SR) are determined
by \beq \label{pislp} \epsilon_1=\fr{1}{p},~\delta_1=-\fr{1}{pq},~
\delta_2=\fr{1+q}{(pq)^2},~\delta_3=-\fr{2q^2+3q+1}{(pq)^3}
,~\delta_4=\fr{6q^3+11q^2+6q+1}{(pq)^4}\eeq which are obtained
from relations in Eq. (\ref{slow-di}) after setting
$\epsilon_1=1/p$\cite{RL}. This inflation goes very well with the
slow-roll expansion. All of H-SR towers are constant for power-law
inflations.
 The
$q$-power spectrum takes the form \beqa
 \label{2ndpssb}
P^{PI,q}_{R_c}(k)& = &\frac{H^4}{(2\pi)^2\dot\phi^2} \Big[ 1
+\fr{1}{p}\Big(4\alpha-\fr{2\alpha}{q}-2\Big) \\ \nonumber
               &+&  \fr{1}{p^2}\Big\{(\fr{2}{q^2}-\fr{8}{q}+8)\alpha^2
-(\fr{2}{q}-4)\alpha+\fr{-8+\pi^2}{2q^2}+\fr{18-2\pi^2}{q}-19+2\pi^2
\Big\}\Big]
\\ \nonumber
       & =&\frac{H^4}{(2\pi)^2\dot\phi^2} \Big[ 1
+\fr{1}{p}\Big(-\fr{1.4592}{q}+0.9184\Big)+\fr{1}{p^2}\Big(\fr{1.99943}{q^2}
-\fr{4.53858}{q} +2.07934\Big)\Big]. \eeqa The $q$-spectral index
can be easily calculated up to third order
 \beq \label{3rdnspi} n^{PI,q}_{s}(k)=1 -4\Big(\fr{1}{p}+\fr{1}{p^2}+\fr{1}{p^3}\Big)\Big(1-\fr{1}{2q}\Big).
\eeq Finally, the $q$-running spectral index is found to be zero
up to $1/p^4$, \beq \frac{d n^{PI,q}_s}{d\ln k}=0. \eeq Even
though  the running spectral index has a complicated from, we find
that for power-law inflations, $\frac{d n^{q}_{s}}{d\ln k}=0$,
irrespective of $q$.
 In the
case of together $q\to 0$ with $p\to \infty$ but $pq \to $ a
finite quantity (equivalently, $\epsilon_1,~\delta_n \to 0$), one
finds de Sitter inflation with $n_s^{PI,q\to 0}=1$. This
corresponds to the extreme slow-roll regime (ESR) with $V$=nearly
constant.

On the other hand, in order to obtain potential slow-roll
parameters (V-SR), we have to choose  explicit potentials which
give rise to power-law inflations (see Appendix B). These are
given  by\cite{Cal1} \beq V^{GB} \sim \phi^6,~~V^{4D} \sim
e^{-\sqrt{2/p}~\phi}, ~~V^{RS-II} \sim \phi^{-2}.\eeq
\begin{table}
 \caption{Power-law inflation potentials and potential slow-roll parameters (V-SR) in patch cosmology.}
 \begin{tabular}{|p{3cm}|p{5cm}|p{7cm}|} \hline
 model   & potential  & V-SR\\ \hline
 GB      & $V^{GB}=\fr{1}{128}\Big(\fr{\kappa^2_5}{16\alpha} \Big)^2\fr{2p-1}{p^4}\phi^6$&
          $\epsilon^V_1=[(2p-1)/2p^4]^{1/3},
~\delta^V_1=-(3/2)[(2p-1)/2p^4]^{1/3},~\delta^V_2=(15/4)[(2p-1)/2p^4]^{2/3},~\delta^V_3=-(105/8)[(2p-1)/2p^4],
~\delta^V_4=(945/16)[(2p-1)/2p^4]^{4/3}$ \\
4D      & $V^{4D}=V_0 \exp(-\sqrt{2\kappa^2_4/p}~\phi)$  &
$\epsilon^V_1=1/p,~\delta^V_1=-1/p,~
\delta^V_2=2/p^2,~\delta^V_3=-6/p^3 ,~\delta^V_4=24/p^4$ \\
 RS-II      &$V^{RS}(\phi)=\fr{4(6p-1)}{3}\fr{\lambda}{\kappa^2_4}\fr{1}{\phi^2}$ & $\epsilon^V_1=6/(6p-1),
~\delta^V_1=-3/(6p-1),~\delta_2^V=27/(6p-1)^2,~\delta_3^V=-405/(6p-1)^3,
~\delta_4^V=8505/(6p-1)^4$ \\ \hline
 \end{tabular}
 \end{table}
 Instead
of an exponential potential $V^{4D}$ for the standard inflation, a
monomial potential of $V^{GB}$ and an inverse power-law potential
of $V^{RS}$ are suitable for power-law inflation. Choosing
coefficients in potentials appropriately, all will take similar
shapes  during  inflation. Their potentials and corresponding
slow-roll parameters appear in TABLE I. For a large $p>1$,
$\epsilon_1 \simeq \epsilon_1^V,~\delta_n \simeq \delta_n^V$  are
found for GB and RS-II cases, whereas one obtains the exact
relations of $\epsilon_1=\epsilon_1^V,~\delta_n=\delta_n^V$ for 4D
case.

 According to WMAP data\cite{Wmap11,Wmap12,Wmap13}, power spectrum normalization
at $k_0=0.05 {\rm Mpc}^{-1}$ is given by ${\cal
A}=0.833^{+0.086}_{-0.083}$ where a normalization factor ${\cal
A}$ is defined by
$P^{ESR}_{R_c}=\frac{H^4}{(2\pi)^2\dot\phi^2}\times {\cal A}
=2.95\times10^{-9}\times {\cal A}$ and scalar spectral index is
$n_s=0.93^{+0.03}_{-0.03}$ at $k_0=0.05 {\rm Mpc}^{-1}$. Running
spectral index is $dn_{s}/d\ln{k}=-0.031^{+0.016}_{-0.018}$ at
$k_0=0.05 {\rm Mpc}^{-1}$ and  tensor-to-scalar ratio at
$k_0=0.002{\rm Mpc}^{-1}$ is $R<0.90$ (95\%CL).

 As is shown in TABLE II,  there exist slightly
 small changes in cosmological parameters. Different potentials give
 slightly different power spectra  and spectral indices
 but give the same running spectral index. Apparently
we find blue (red) power spectrum corrections to RS-II (GB,4D)
inflations (see Fig. 2). We note that this is not a crucial result
because we measure only a normalization factor ${\cal A}$ of the
power spectrum from the WMAP. Also we have red spectral indices
for all cases (see Fig. 3).

\begin{table}
 \caption{Power spectrum normalization $\Big({\cal A}^{PI}\Big)$,
  spectral index $\Big(n_s^{PI}\Big)$, and running spectral index $\Big(\frac{d n^{PI}_s}{d\ln k}\Big)$
   based on H-SR towers.
 $^{\rm a}$ Here $\approx 0$ means $-7.10543\times 10^{-15}/p^3+
8.52651 \times 10^{-14}/p^4\approx 0$.}
\begin{tabular}{|c|c|c|c|} \hline
  model & ${\cal A}^{PI}$ & $n_s^{PI}$
   & $\frac{d n^{PI}_s}{d\ln k}$\\
  \hline
  GB & $ 1
-\fr{1.2740}{p}-\fr{0.229741}{p^2}$ &
$1-\fr{1}{p}-\fr{1}{p^2}-\fr{1}{p^3}$
& $\approx 0~^{\rm a}$ \\
4D &$1 -\fr{0.540726}{p}-\fr{0.459731}{p^2}$
&$1-\fr{2}{p}-\fr{2}{p^2}-\fr{2}{p^3}$ &0 \\
  RS-II &$ 1
+\fr{0.1888}{p}+\fr{0.309928}{p^2}$
&$1-\fr{3}{p}-\fr{3}{p^2}-\fr{3}{p^3}$& 0 \\ \hline

\end{tabular}
 \end{table}

As a guideline to the power-law inflation, choosing $p=101$
\footnote{In Ref.\cite{RL}, the authors choose $p=101$ to take
$N=50$ $e$-foldings before the end of inflation. However, they use
a monomial potential of $V=\phi^a,~a=2,4,6$ which give rise to
chaotic large-field inflation, to obtain corrections to the power
spectrum for 4D and RS-II cases. Here our comparison test is based
on the power-law inflation with different potentials depending on
GB, 4D, and RS-II. Also, a choice of $p=101$ satisfies
$2p>>1,~6p>>1$, which leads to $\epsilon_1\simeq
\epsilon_1^V,~\delta_n \simeq \delta_n^V$. That is, there is no
sizable difference between H-SR and V-SR towers.} leads to ${\cal
A}^{PI,RS-II}=1,~1.00187,~1.0019$ for zero, first, second order
corrections, respectively, whereas ${\cal
A}^{PI,4D}=1,~0.994646,~0.99461$ and ${\cal A}^{PI,GB}=1,~0.987386
(0.987407),~0.987364 (0.987385)$. Also we find that
$n_s^{PI,RS-II}=0.970297 (0.970248),~0.970003 (0.969953),~0.97
(0.96995)$ for zero, first, second-order corrections,
 while $n_s^{PI,4D}=0.980198,~0.980002,~0.98$ and
$n_s^{PI,GB}=0.990099 (0.990115),~0.990001 (0.990018),~0.99
(0.990017)$. Here ($\cdots$) are calculated using V-SR towers (see
TABLE III). In the case of $q\to \infty$, we have ${\cal
A}^{PI,\infty}=1,~1.00909,~1.0093$ and
$n_s^{PI,\infty}=0.960396,~0.960004,~0.96$. Although its potential
is not yet known,  this case provides us the smallest spectral
index and the largest power spectrum.  We find from the above that
first and second-order corrections lie within the uncertainty.
Fitting of $n_s$ to the WMAP seems to be beyond the uncertainty
for a $p=101$ case. According to Ref.\cite{LL}, however, a
constraint on 4D power-law inflation is given by
$0<\epsilon_1<0.019$ and $p>53$. Here we choose an appropriate $p$
between $50<p<110$ to fit the data within the uncertainty. In the
case of GB power-law inflation, we may loosen the lower bound of
$p$ to fit the data.

 Since recent observations including WMAP have
restricted viable inflation models to regions close to the
slow-roll limit, our second-order corrections to the patch
cosmology are rather small. If one uses V-SR towers with
$50<p<110(2p-1\simeq 2p,~6p-1\simeq 6p)$, also we lead to the same
conclusion (see TABLE III).  Hence we confirm that in the
slow-roll limit, observations of the primordial perturbation
spectra cannot distinguish  between GB, RS-II, and 4D power-law
inflations\cite{LT}.  Hence we need to introduce the tensor
spectrum, especially for the tensor-to-scalar ratio.

 \begin{table}
 \caption{Power spectrum, spectral index, and running spectral index based on V-SR
 towers. $^{\rm b}$ Here $\approx 0$ means $-7.10543\times
10^{-15}\Big[(2p-1)/(2p^4)\Big]+ 8.52651 \times 10^{-14}
\Big[(2p-1)/(2p^4)\Big]^{4/3}\approx 0$.}
\begin{tabular}{|c|c|c|c|}
 \hline
  model & ${\cal A}^{PI}$ & $n_s^{PI}$
   &$\frac{d n^{PI}_s}{d\ln k}$ \\
  \hline
  GB & $ 1
-1.2740\Big[\fr{2p-1}{2p^4}\Big]^{\fr{1}{3}}-0.229741\Big[\fr{2p-1}{2p^4}\Big]^{\fr{2}{3}}$
&
$1-\Big[\fr{2p-1}{2p^4}\Big]^{\fr{1}{3}}-\Big[\fr{2p-1}{2p^4}\Big]^{\fr{2}{3}}-\Big[\fr{2p-1}{2p^4}\Big]$
& $\approx 0~^{\rm b}$ \\
4D &$1 -\fr{0.540726}{p}-\fr{0.459731}{p^2}$
&$1-\fr{2}{p}-\fr{2}{p^2}-\fr{2}{p^3}$ &0 \\
  RS-II &$ 1
+0.1888\Big[\fr{6}{6p-1}\Big]+0.309928\Big[\fr{6}{6p-1}\Big]^{2}$
&$1-3\Big[\fr{6}{6p-1}\Big]-3\Big[\fr{6}{6p-1}\Big]^{2}-3\Big[\fr{6}{6p-1}\Big]^{3}$&
0
\\ \hline

 \end{tabular}
 \end{table}

\begin{figure}[t!]
\includegraphics{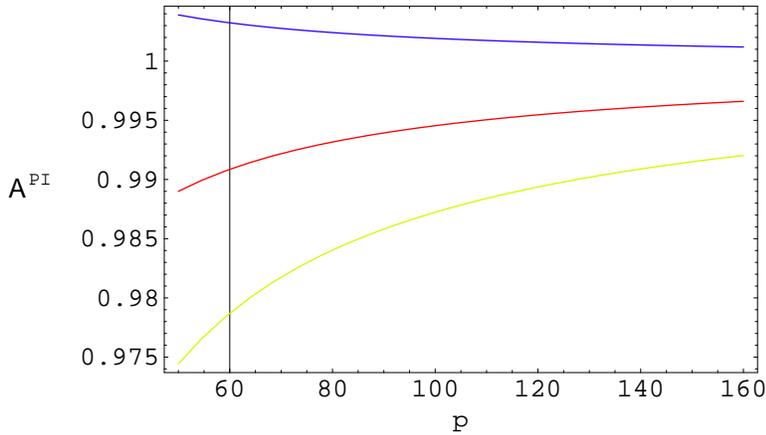}
\caption{Plot of the power spectrum normalization ${\cal A}^{PI}$
for power-law inflation with $a(t)\sim t^p$.  From the top curve
to the bottom one, one finds RS-II(blue), 4D(red), and GB(yellow),
respectively. An appropriate value $p$ is between $p=50$ and
$p=110$, and a line of $p=60$ is introduced for comparison.}
\label{fig2}
\end{figure}
\begin{figure}[t!]
\includegraphics{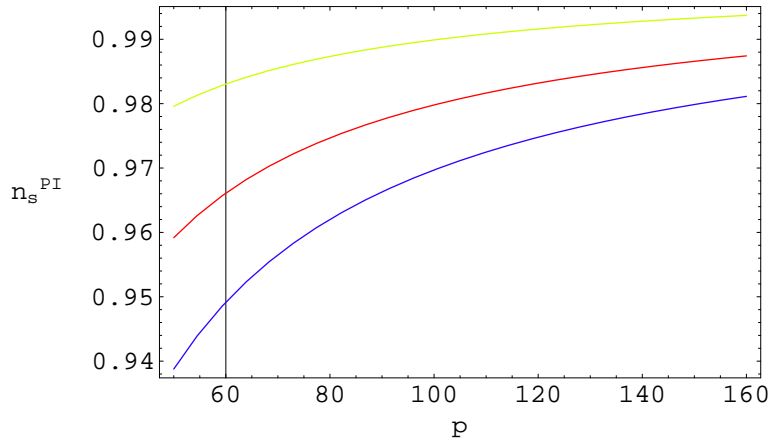}
\caption{Plot of  the spectral index $n_s^{PI}$ for power-law
inflation with $a(t)\sim t^p$. From the top curve to the bottom
one, one finds GB(yellow), 4D(red), and RS-II(blue), respectively.
An appropriate value $p$ is between $p=50$ and $p=110$, and a line
of $p=60$ is introduced for comparison.} \label{fig3}
\end{figure}

 \section{Consistency relation in patch cosmology}

The tensor-to-scalar ratio $R$ is defined by \beq
R=16\fr{A_{T,q}^2}{A_{S,q}^2}.\eeq Here the $q$-scalar amplitude
to zero order is
 given by \beq
A_{S,q}^2=\fr{4}{25}P_{R_c}^{q,ESR} \eeq
 with the extreme slow-roll power spectrum \beq
 P_{R_c}^{q,ESR}=\fr{3q\beta^{2-\theta}_q}{(2\pi)^2}\fr{H^{2+\theta}}{2\epsilon_1}=
 \fr{1}{(2\pi)^2}\fr{H^4}{\dot{\phi}^2}.
 \eeq
The 4D($q=1$) tensor amplitude to zero order is given by \beq
A_{T,4D}^2=\fr{1}{50}P_T^{ESR} \eeq
 with
$P_T^{ESR}=(2\kappa_4)^2\Big(\fr{H}{2\pi}\Big)^2$ because a tensor
can be expressed in terms of two scalars like $\delta \phi$ with a
factor $2\kappa_4$.  On the other hand, the tensor spectra for
GB($q=2/3$) and RS-II($q=2$) are known only for de Sitter
brane\cite{LMW,DLMS}. These are given by \beq
A_{T,q}^2=A_{T,4D}^2F_{\alpha}^2(H/\mu), \eeq where \beq
F_{\beta}^{-2}(x)=\sqrt{1+x^2}-\Big(\fr{1-\beta} {1+\beta}\Big)x^2
\sinh^{-1}\Big(\fr{1}{x}\Big).\eeq  In three regimes, we
approximate $F_{\beta}^2$ as $F_q^2$: $F_1^2=1\approx
F_{\beta}^2(H/\mu \ll 1)$ for 4D case; $F_2^2=3H/(2\mu)\approx
F_{\beta=0}^2(H/\mu \gg 1)$ for RS-II case;
$F_{2/3}^2=(1+\beta)/(2\beta)(\mu/ H) \approx F_{\beta}^2(H/\mu
\gg 1)$ for GB case. The tensor amplitude to zero order is given
by \beq
 A_{T,q}^2=\fr{3q\beta^{2-\theta}_q}{(5\pi)^2}\fr{H^{2+\theta}}{2\zeta_q(h)}
 \eeq
 with $\zeta_1(h)=\zeta_{2/3}(h)=1$ and $\zeta_2(h)=2/3$\cite{Cal3}.
 Then the tensor-to-scalar ratio is determined by
\beq
R_q=16\fr{A_{T,q}^2}{A_{S,q}^2}=16\fr{\epsilon_1}{\zeta_q(h)}.\eeq
Considering $n_{T,1}=-2\epsilon_1,~n_{T,2/3}=-\epsilon$ and
$n_{T,2}=-3\epsilon_1$, one finds that \beq
R_1=-8n_{T,1}=16\epsilon_1,~R_2=-8n_{T,2}=24\epsilon_1,~R_{2/3}=-16n_{T,2/3}=16\epsilon_1.\eeq
The above shows that the RS-II consistency relation is the same
for that of 4D case  but the GB consistency relation is different
from RS-II and 4D cases.

In the de Sitter brane  approach with $a\sim e^{Ht}$, we have no
non-zero H-SR towers. The zero-order scalar amplitude for GB
braneworld is given by $A_{S,dS}^2=A_{S,4D}^2G_{\beta}^2(H/\mu)$
where $G_{\beta}^2(x)
=\Big[\fr{3(1+\beta)x^2}{2\sqrt{1+x^2}(3-\beta+2\beta
x^2)+2(\beta-3)}\Big]^3$. In the 4D limit, we have
$G_{\beta}^2(H/\mu \ll 1) \approx 1$. In the GB regime,
$G_{\beta}^2(H/\mu \gg 1) \approx
\fr{27}{64}\Big(\fr{1+\beta}{\beta}\Big)^3\fr{\mu^3}{H^3}$, while
in the RS-II regime, $G_{\beta=0}^2(H/\mu\gg 1)\approx
\fr{H}{2\mu}$. These lead to
$A_{S,q}^2=\fr{3q\beta^{2-\theta}_q}{(5\pi)^2}\fr{H^{2+\theta}}{2\epsilon_1}$.
In the extreme slow-roll regime of $\epsilon_1,~\delta_n \to 0$,
one finds the same amplitude of
$\fr{1}{(2\pi)^2}\fr{H^4}{\dot{\phi}^2}$,  as found in the de
Sitter brane approach. However, the de Sitter picture is basically
different from ours because we work with slow-roll approximation
of $\epsilon_1<\xi,|\delta_n|<\xi^n$ for $\xi<1$, but not with a
case with $\epsilon_1,\delta_n \to 0$ for de Sitter
brane\cite{ACMZ}. In other words, we work with
$\rho+p=\dot{\phi}^2$ but not a case :$\rho+p \to 0 \Rightarrow
V=$ constant    as in de Sitter brane. In the de Sitter brane
approach,  we cannot make any slow-roll approximation because  de
Sitter space means that $H$= constant during inflation. In this
sense the slow-roll approximation to GB braneworld based on  de
Sitter brane  to obtain a tensor spectrum leads to  an obscure
computation.

\section{Discussions}
Our second-order corrections which appear even slightly different
from those of the standard inflation, could not play a role in
distinguishing between GB, RS-II, 4D slow-roll inflations. Thus it
is necessary to introduce the tensor power spectrum to distinguish
them. The  reason is as follows. These models are based on the
same perturbation scheme given by the Mukhanov equation
(\ref{eqsn}) with slightly different potential-like terms:
$\frac{1}{z}\frac{d^2z}{d\tau^2}$. This patch cosmology with an
inflaton gives us similar results in the slow-roll expansion
except a relation of $\dot \epsilon_1
=2H(\epsilon_1^2/q+\epsilon_1\delta_1)$ which affects second-order
and more higher-orders only. For three different potentials,  we
find the nearly same power-law inflation. In the slow-roll limit,
these give us the nearly same cosmological parameters. Since an
introduction of a Gauss-Bonnet term in the braneworld could not
distinguish between GB, 4D, and  RS-II, we need to introduce the
tensor spectrum. Thus the observation of gravitational waves may
be helpful to select  a desired inflation model.

Even though there exist a $q$-dependent term of $-8\epsilon_1^2/q$
in the lowest-order of the running spectral index, we find that $
\frac{d n^{q}_{s}}{d\ln k}=0$, irrespective of $q$, when choosing
power-law potentials. This shows the nature of power-law inflation
in the patch cosmology.  It compares with the WMAP data of
$dn_{s}/d\ln{k}=-0.031^{+0.016}_{-0.018}$ at $k_0=0.05 {\rm
Mpc}^{-1}$.

We have a few of comments on other cases in patch cosmology.
  From Eq.(\ref{3rdnspi}), for  $q \to \infty$, one finds an interesting case of
$n_s^{PI,\infty} \to 1-4(1/p+1/p^2)$.  Also we find a
scale-invariant spectral index of $n_s=1$ for $q=1/2$,
irrespective of $p$. Although we don't know the corresponding
model explicitly, it will be  located beyond the GB high-energy
regime. In the case of $q>1/2$, we have a red spectral index,
whereas for $q<1/2$, we have a blue index. In the limit of
$q\to0$, unfortunately one finds a largely blue spectral index of
$n_s^{PI,q\to0}> 1$ which is ruled out from the data. Hence an
appropriate region to a patch parameter $q$ is given by $1/2\le
q<\infty$ which provides  a restriction : $1-4(1/p+1/p^2) <n_s \le
1$.

Finally, we emphasize that our calculation based on the Mukhanov
equation is reliable up first order corrections. At this stage we
don't know whether or not the second order corrections are smaller
than the effect of 5D metric perturbations. Even though we
calculate cosmological parameters up to second order to understand
the power-law nature of patch cosmology,  second order corrections
are less important because these are rather small than  first
order corrections in the slow-roll limit and these are not yet
proved to be reliable.

\subsection*{Acknowledgements}
We thank  Hungsoo Kim, H. W. Lee and G. Calcagni for helpful
discussions.  K. Kim was in part supported by KOSEF, Astrophysical
Research Center for the Structure and Evolution of the Cosmos. Y.
Myung was supported by the Korea Research Foundation Grant
(KRF-2005-013-C00018).

.
\newpage
\subsection*{Appendix A: Derivation of the Mukhanov equation
including 5D metric perturbations.} We start with Eq.(C.5) in
Ref.\cite{KS} expressed in terms of $Q= \delta
\phi-\dot{\phi}\psi/H$,
\begin{equation} \label{A1}
 \ddot{Q} +3H \dot{Q} +\frac{k^2}{a^2}Q +
\Big[\frac{\ddot{H}}{H}-2\frac{\dot{H} V'}{H
\dot{\phi}}-2\Big(\frac{\dot{H}}{H}\Big)^2+V''\Big]Q=J.
\end{equation}
Here $J$ is  the contribution from the 5D metric perturbations
given by
\begin{eqnarray}
J=\frac{\dot{\phi}}{H}\Big[\Big(\frac{2\ddot{H}}{\dot{H}}-\frac{\dot{H}}{H}\Big)\Delta
v +\frac{k^2}{3a^2}A +
\frac{k^2}{3a^2}\Big(1-\frac{\dot{H}}{\alpha^2_1}\Big)\psi \nonumber \\
+\frac{k^4}{9a^4} \Big(2-\frac{\dot{H}}{\alpha^2_1} \Big) \int dm
E(m,k)l^2Z_0(ml/a)e^{-i\omega T(t)} \Big],
\end{eqnarray}
where the detailed information on the unknown functions ($\Delta
v,\alpha_1^2, E(m,k), \cdots$) are given by  Ref.\cite{KS}. We
note that the above equation is derived by using the braneworld
scenario without the Gauss-Bonnet term. In this work we are
interested in its patch approximation.  We wish to derive the
corresponding Mukhanov equation including the effect of 5D metric
perturbations. Using $Q^q=u^q/a$, Eq.(\ref{seq}) and its
derivative, and Eq.(\ref{slpa}), we obtain the following equation
from Eq.(\ref{A1}) exactly:
\begin{equation}
\label{A3} \frac{d^2u^q_k}{d\tau^2} +
\left(k^2-\frac{1}{z}\frac{d^2z}{d\tau^2}\right)u^q_k =
a^3J^q\end{equation} with the $q$-dependent potential
$\frac{1}{z}\frac{d^2z}{d\tau^2}$ in Eq.(\ref{z-pot}) and patch
approximation $J^q$ to $J$. In the limit of $J^q\to 0$, we recover
the Mukhanov equation (\ref{eqsn}). Koyama and Soda showed
implicitly that $J \to 0$ could be achieved on super-horizon
scale\cite{KS}. According to Koyama {\it et al}\cite{KLMW}, it
turns out that  $J^{q=1}$ at low-energy of $H/\mu\ll 1$ and
$J^{q=2}$ at high-energy of $H/\mu \gg 1$ with $\beta=0$ are
smaller than first order corrections to on super-horizon scale.
Similarly, we expect that $J^{q=2/3}$ at high-energy of $H/\mu \gg
1$ with $\beta\not=0$ is smaller than first order corrections. At
this stage, we don't know whether or not  $J^q$ is  smaller than
second order corrections.  At first order of the slow-roll
expansion, the ratio $J^{q=2}$ to  first-order term takes the form
of $J^{q=2}/\dot{H} Q \sim \frac{k^4}{(aH)^4}$ at high energies
and thus it goes to zero on super-horizon scale of $k \ll aH$.  On
the other hand, the ratio $J^{q=2}$ to second-order term takes the
form of $J^{q=2}H^2/\dot{H}^2 Q \sim
\frac{1}{\epsilon_1}\frac{k^4}{(aH)^4}$ at high energies. If $k$
is enough large than $aH$ on super-horizon scale, it seems  that
the second order calculation is reliable.  However, this does not
show that the effect of 5D metric perturbations is less definitely
than second order corrections. On the other hand, the effect of 5D
metric perturbations is less certainly than  first order
corrections.

\newpage
\subsection*{Appendix B: Potential slow-roll parameters in the patch cosmology}
The potential  slow-roll parameters (V-SR)  are given  by
\begin{equation}
\epsilon^V_1=\frac{q}{6\beta^2_q}\frac{V'^2}{V^{1+q}},~~
\delta^V_1=-\frac{1}{3\beta^2_q}\Big[\frac{V''}{V^q}-\frac{V'^2}{V^{1+q}}\Big],
\end{equation}
\begin{eqnarray}
\delta_2^V =
(-\frac{1}{3\beta^2_q})^2&\Big[&\frac{V'''V'}{V^{2q}}+
\frac{V''^{2}}{V^{2q}}-\frac{5q}{2}\frac{V''V'^{2}}{V^{2q+1}}
+\frac{q}{2}(q+1)\frac{V'^{4}}{V^{2q+2}}\Big]            ,\\
\delta_3^V=(-\frac{1}{3\beta^2_q})^3
&\Big[&\frac{V''''V'^{2}}{V^{3q}}-4q\frac{V'''V'^{3}}{V^{3q+1}}
+9q\frac{V''^{2}V'^{2}}{V^{3q+1}}+5q(5q+1)\frac{V''V'^{4}}{V^{3q+2}}\\
\nonumber
&+&4\frac{V'''V''V'}{V^{3q}}+\frac{V''^{3}}{V^{3q}}-\frac{q}{2}(q+1)(\frac{3q}{2}+2)
\frac{V'^{6}}{V^{3q+3}}\Big]  ,\\
\delta_4^V=(-\frac{1}{3\beta^2_q})^4&
\Big[&\frac{V'''''V'^{3}}{V^{4q}}+7\frac{V''''V''V'^{2}}{V^{4q}}
-6q\frac{V''''V'^{4}}{V^{4q+1}}-42q\frac{V'''V''V'^{3}}{V^{4q+1}}\\
\nonumber
 &+&q(\frac{57q}{4}+9)\frac{V'''V'^{5}}{V^{4q+2}}-29q\frac{V''^{3}V'^{2}}{V^{4q+1}}
  +q(\frac{197q}{4}+34)\frac{V''^{2}V'^{4}}{V^{4q+2}}\\
   \nonumber
    &-&q(\fr{71}{4}q^{2}+\frac{139}{4}q+17)
     \frac{V''V'^{6}}{V^{4q+3}}
     +11\frac{V'''V''^{2}V'}{V^{4q}}+\frac{V''^{4}}{V^{4q}}\\
      \nonumber
       &+&4\frac{V'''^{2}V'^{2}}{V^{4q}}+
        \frac{q}{2}(q+1)(\frac{3q}{2}+2)(2q+3)\frac{V'^{8}}{V^{4q+4}}\Big].\\
          \nonumber
\end{eqnarray}
Here the prime($\prime$) denotes the derivative with respect to
$\phi$.

\end{document}